\documentclass[a4paper,twocolumn]{article} 
                          
\usepackage{graphicx}   
\usepackage{epsfig} 
\usepackage{epstopdf}
\usepackage{times} 
\usepackage{amsmath} 
\usepackage{amssymb}  
\usepackage{theorem}
\usepackage[]{units}
\usepackage{subfig}
\usepackage{psfrag}
\usepackage{wrapfig}
\usepackage{color}
\usepackage[percent]{overpic}

\usepackage[dvipsnames]{xcolor}

\usepackage[]{algorithm}
\usepackage[]{algpseudocode}

\usepackage{multirow}

\usepackage[font={footnotesize}]{caption} 

\title{
Modeling pressure pulsation and backflow in progressing cavity pumps with deformable stator
}

\author{Jens M\"uller, Yashar Kouhi, Sebastian Leonow, Martin M\"onnigmann\thanks{Corresponding author.}\\
Automatic Control and Systems Theory, Department of Mechanical Engineering,\\
	Ruhr-Universit\"at Bochum, 44801 Bochum, Germany.\\ \small E-mail: \tt \{jens.mueller-r55, yashar.kouhi, \\ 
	\small \tt sebastian.leonow, martin.moennigmann\}@rub.de}

\begin{document}

\maketitle

\begin{abstract}
This contribution studies the impact of the rotor-stator interaction in a single-stage progressing cavity pump on the flow rate and pressure. Specifically, we investigate the effect of the rotor movement on the sealings formed with deformable stators for various speeds and pressures. Sealings are reconstructed with the help of a geometric 3D model. We analyze the tangential and radial deviation of the rotor from its reference path and show that the radial deviation affects the flow rate, whereas the tangential deviation affects the pressure dynamics. The conjectures are confirmed with a laboratory test setup. 

$\copyright$ 2021. This manuscript version is made available under the CC-BY-NC-ND 4.0 license http://creativecommons.org/licenses/by-nc-nd/4.0/
\end{abstract}

\section{Introduction} \label{intro}
Progressing cavity pumps (PCPs) belong to the group of positive displacement pumps. They are used in industrial applications for a wide variety of media. The volumetric flow rate of PCPs is known to be an almost linear function of the rotational speed \cite{cholet_2013}.
As the differential pressure increases, a growing amount of fluid flows against the conveying direction. 
The volumetric efficiency of the pump obviously decreases with growing backflow. An accurate model for the backflow is required to predict the flow rate and the volumetric efficiency of the pump. Furthermore, an understanding of the physical processes that cause backflow is advantageous, since results can be transferred to other pump geometries and can be used for the examination of the correlation between backflow and wear. 

Methods for determining the backflow in a PCP are, for example,  given in \cite{gamboa_2003}, \cite{nguyen_2016}, and \cite{pessoa_2009}. However, these approaches are conservative in that they assume the rotor to move on an ideal path inside the stator. Consequently, the dimensions of the contact area (respectively channel geometry for clearance fit) between rotor and stator are calculated for the ideal case. This assumption seems to be sufficient for PCPs with metallic stators. Previous results in \cite{mueller_2019} and \cite{mueller_2017}, however, reveal that neglecting any deviation of the ideal path can lead to incorrect results for PCPs with elastomer stators.

The investigations in \cite{wirth_1993} on PCPs with elastomer stators show there exists a correlation of the movement of the rotor to the shape of the sealings inside the pump. However, the authors of \cite{wirth_1993} claim that the shape of the sealings cannot be described quantitatively. In \cite{belcher_1991} and \cite{mueller_2017} tactile and inductive sensors, respectively, were used to investigate the motion of the rotor. Since the movement of the rotor as a function of the rotational angle was not investigated, the effect of the rotor movement on the sealings could not be analyzed.

It is the aim of this contribution to model the pressure pulsation and backflow of PCPs as a function of the actual rotor movement. To this end, we study the relation between the actual movement of the rotor and the shape of the sealings inside the pump for a single-stage PCP with elastomer stator. We distinguish two components of the deviation of the rotor from its ideal path from one another and show that one component governs the amount of backflow, whereas the other component governs the pressure dynamics. Both claims are verified experimentally.

Section~\ref{sec2} introduces the pump geometry, the basic interaction between rotor and stator, and preliminaries necessary for the analysis of the load case. Section~\ref{sec3} states the main results concerning the modeling of backflow and pressure pulsation. Section~\ref{sec4} presents the laboratory test setup and verifies the given conjectures with measurement data. A conclusion is given in Section~\ref{sec5}.

\section{Progressing cavity pump} \label{sec2}
The pump under consideration is a single-stage, single lobe progressing cavity pump without significant wear. It consists of a helical steel rotor and a deformable elastomer stator. The inner geometry of the stator is a twisted slot hole with the stator pitch $P_S$. The lengthwise orientation of the stator (set during the pump assembly) is specified by the stator orientation angle $\theta$. The rotor is assumed to be rigid.

\subsection{Rotor movement for the no-load case} \label{sec21}
\begin{figure}
\begin{center}
\includegraphics[width=7.4cm]{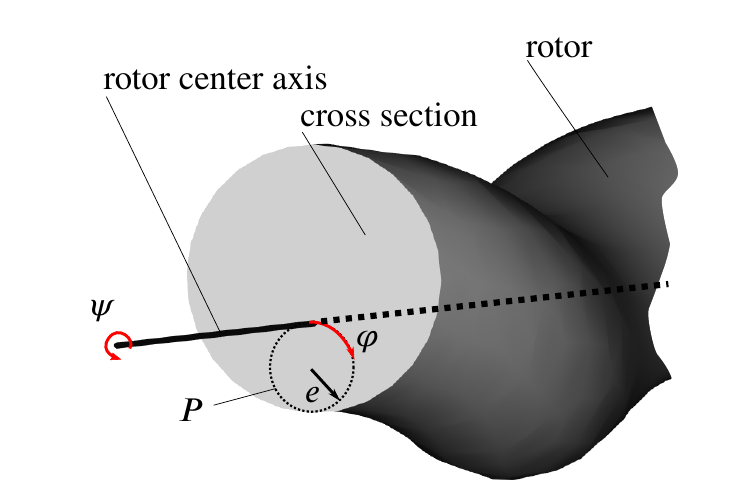}    
\caption{Superposed rotor movements with corresponding angles $\psi$ and $\varphi$ for the ideal case.} 
\label{fig:rotor_coordinates}
\end{center}
\end{figure}

The rotor performs two superposed movements: the rotation around its center axis (angle $\psi$) and the movement of the center axis on a closed curve (angle $\varphi$). Figure~\ref{fig:rotor_coordinates} illustrates the coordinates for both movements for the ideal case. The closed curve in Figure~\ref{fig:rotor_coordinates} is referred to as the \textit{path} $P$ of the rotor. For ideal geometries and in the absence of disturbances, the path is a circle with radius $e$.  

\subsection{Rotor-stator interaction} \label{sec22}
The rotor-stator contact areas perform two tasks: they act as a radial bearing for the rotor and thus force the rotor to stay on, or close to, the ideal path $P$. At the same time, the contact areas build up sealings inside the pump that separate cavities. We analyze the interactions of the cross section of the rotor and the stator to investigate the shape of the sealings in the present section. 

\begin{figure}
\begin{center}
\includegraphics[width=5.5cm]{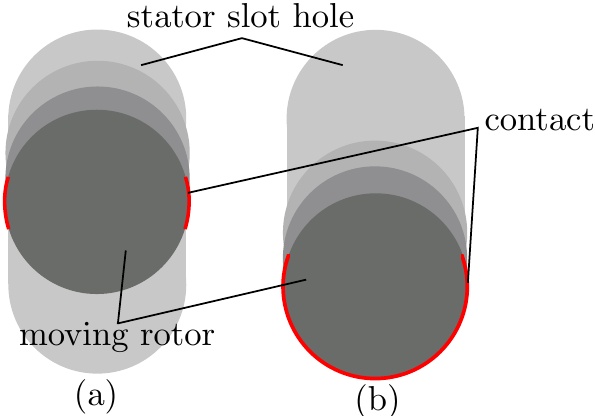}    
\caption{Planar rotor-stator contact.} 
\label{fig:contact_situations}
\end{center}
\end{figure}

Two types of rotor-stator contact are sketched in Figure~\ref{fig:contact_situations}. Any cross section of the rotor is either located between the straight walls of the slot hole (a) or located at one end of the slot hole (b). 

\begin{figure}
\begin{center}
\includegraphics[width=\columnwidth]{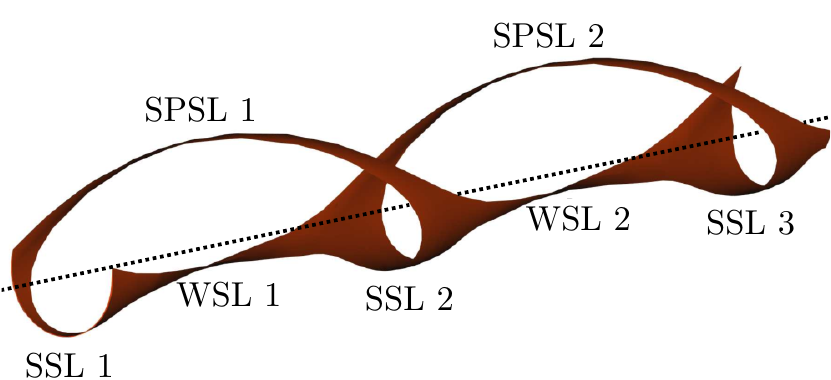}   
\caption{Ideal sealings inside a single-stage progressing cavity pump for a fixed $\varphi$.} 
\label{fig:sealings_ideal}
\end{center}
\end{figure}

These two contact types give rise to three types of sealings~\cite{pan_2015}. The two areas highlighted in red color in Figure \ref{fig:contact_situations} (a) result in the spiral seal line (SPSL) and the warping seal line (WSL), which are shown in Figure \ref{fig:sealings_ideal}. Due to the helical geometry of the rotor, the WSL and the SPSL do not have the same properties and must be distinguished from one another. The area highlighted in Figure \ref{fig:contact_situations} (b) results in the semicircle seal line (SSL), which is also shown in Figure \ref{fig:sealings_ideal}. 

The three distinct types of sealings are connected to each other and form a continuum in the case of an ideal placement of the rotor.

According to \cite{gamboa_2003}, \cite{wirth_1993}, and \cite{zheng_2018} two flow components determine the amount of  backflow through the sealings. The first component arises due to the relative velocity between rotor and stator. The differential pressure across the sealings causes the second component. According to previous analyses in \cite{pessoa_2009} and our analysis of the characteristic curve \mbox{($Q$/$n$-curve)}, the influence of the first component is small, so that we neglect it. The pressure distribution inside the pump needs to be known for the investigation of the second component.

\begin{figure}
\begin{center}
\includegraphics[width=\columnwidth]{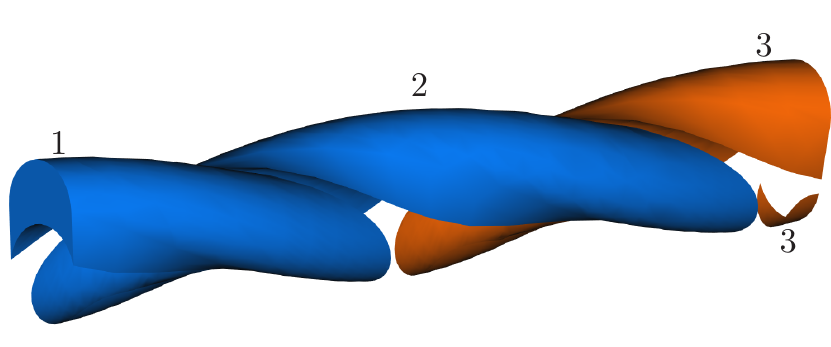}    
\caption{Cavities inside a single-stage progressing cavity pump for a fixed $\varphi$.} 
\label{fig:volume_ideal}
\end{center}
\end{figure}

Let $p_\textrm{p}$ and $p_\textrm{s}$ refer to the pressure on the pressure and suction side, respectively. A cavity open to  the pressure or the suction side attains the associated pressure. The cavity encapsulated inside the pump is assumed to attain the suction side pressure \cite{mueller_2019}. A sketch of a typical situation is shown in Figure~\ref{fig:volume_ideal}, where orange and blue colors indicate $p_\textrm{p}$ and $p_\textrm{s}$, respectively. The cavities labeled $3$ are connected since they are both open to the pressure side. Cavity $1$ is open to the suction side, whereas cavity $2$ is closed to both sides.

The differential pressure across the pump is defined as 
\begin{equation} \label{eq:delta_p}
\Delta p= p_{\textrm{p}}-p_{\textrm{s}} \, .
\end{equation}
A pressure difference $\Delta p \not= 0$ across a sealing is a necessary condition for backflow.  
It is evident from Figure~\ref{fig:volume_ideal} that $\Delta p \not= 0$ holds for four of the sealings, 
specifically for SSL 2, WSL 2,  SPSL 2, and SSL 3 shown in Figure~\ref{fig:sealings_ideal}. 
The differential pressure is zero (indicated by matching colors) for the remaining sealings, leading to a vanishing backflow. 

Backflow only occurs if one or more of the sealing degenerate. This may happen because of wear, or because the rotor deviates from its ideal position due to differential pressure.
We assume the pump is not significantly worn and manufactured in a way that no sealing degenerates for $\Delta p= 0$. 
We outline the effects of $\Delta p\ne 0$ 
on the rotor position in Sections \ref{sec23} and \ref{sec24}.

\begin{figure}
\begin{center}
\includegraphics[trim= 0 0 0 0,clip,width=\columnwidth]{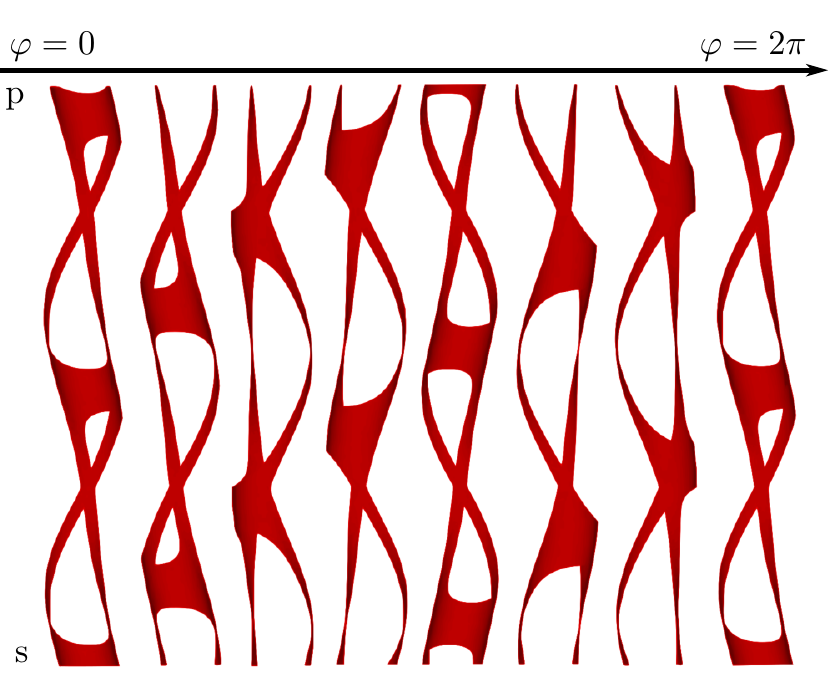}    
\caption{Location of ideal sealings as a function of $\varphi$. Suction side and pressure side indicated by s and p.} 
\label{fig:sealing_movement}
\end{center}
\end{figure}

All statements made so far apply to the locations of the ideal sealings and cavities for an arbitrary but fixed value of $\varphi$. As the pump is operated, the angle $\varphi$ changes and causes the sealings to move 
from the suction side to the pressure side. Simultaneously, the sealings and cavities perform a rotational movement. Both movements are illustrated in Figure~\ref{fig:sealing_movement}. 

\subsection{Geometric sealing model} \label{sec23}
We introduce a geometric 3D model to reconstruct the sealings inside the pump for the load case.
Let $\mathcal{R}$ and $\mathcal{S}$ refer to sets of points $(x,y,z)$ that represent the rotor and stator geometry, respectively. The rotor-stator interaction can then be investigated by analyzing the intersection
\begin{equation} \label{eq:rotor_tilt_equation}
\mathcal{C} = \mathcal{R} \cap \mathcal{S} \  .
\end{equation}

\begin{figure}
\begin{center}
\includegraphics[width=\columnwidth]{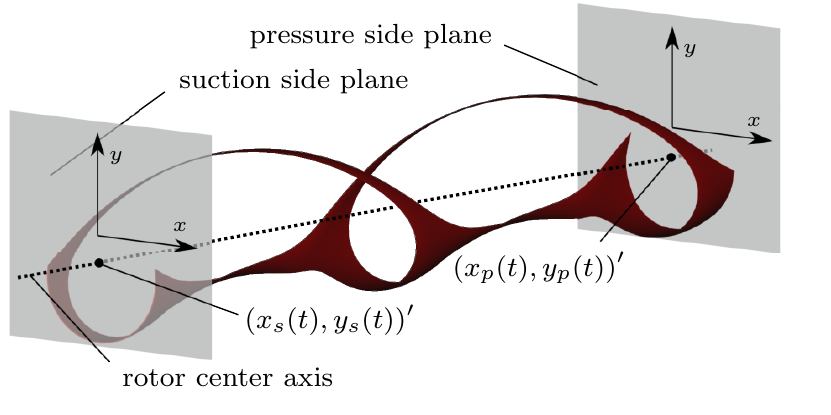}   
\caption{Coordinates of the rotor position.} 
\label{fig:rad_tan_coordinates}
\end{center}
\end{figure}

The intersection $\mathcal{C}$ equals the shape of the sealings formed by the rotor-stator contact.
Figure~\ref{fig:rad_tan_coordinates} depicts $\mathcal{C}$ and the intersection of the rotor center axis with the suction side plane and the pressure side plane. 
The coordinates $\left(x_s(t), y_s(t)\right)'$ denote the intersection of the rotor center axis and the suction side plane and thus represent the position of the rotor end at the suction side. The coordinates $\left(x_p(t), y_p(t)\right)'$ refer to the intersection of the rotor center axis with the pressure side plane. The angle $\psi$ and the coordinates $\left(x_s(t), y_s(t)\right)'$ and $\left(x_p(t), y_p(t)\right)'$ uniquely define the position of $\mathcal{R}$, i.e., the position of the rigid rotor inside the stator. 

The correct representation of the sealings $\mathcal{C}$ by \eqref{eq:rotor_tilt_equation} depends on the exact representation of rotor and stator in $\mathcal{R}$ and $\mathcal{S}$, respectively. Due to manufacturing tolerances, the real dimensions of rotor and stator typically differ from the available data, e.g. the computer aided manufacturing data, in particular in the case of an elastomeric stator. We compensate for this mismatch with an effective rotor diameter 
\begin{equation} \label{eq:delta_d}
 d_{\mathcal{R},\text{eff}} =Kd_{\mathcal{R}} \, ,
\end{equation}
where $d_{\mathcal{R}}$ is the rotor diameter according to the manufacturing data and $K$ is a correction factor. In this way, we ensure the geometric 3D model to resemble the real pump.

The evolution of the sealings over time with rotating rotor can be analyzed with time-series of $\psi$, $\left(x_s(t), y_s(t)\right)'$ and $\left(x_p(t), y_p(t)\right)'$, and by determining $\mathcal{C}$ by \eqref{eq:rotor_tilt_equation} for every sample of the time-series. If the rotor moves on its ideal path inside the stator, the reconstruction of the sealings with the help of the geometric 3D model results in the time series shown in Figure~\ref{fig:sealing_movement}. 
Results obtained under load conditions are shown in Figure~\ref{fig:SSL_quality} and discussed in Section~\ref{sec4}. 

\subsection{Load dependence of the rotor path}\label{sec24}
\begin{figure}
\begin{center}
\includegraphics[trim={0 0 30 20},width=5cm]{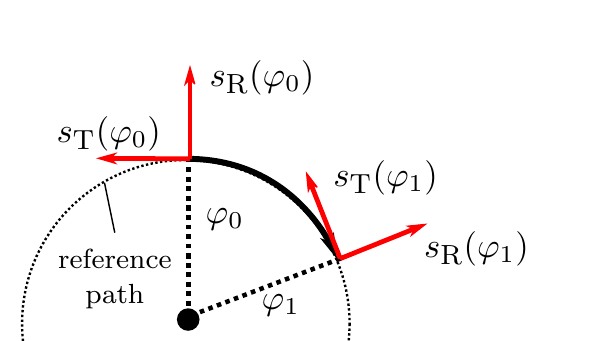}    
\caption{RT-coordinate system for two rotational angles $\varphi_0$ and $\varphi_1$.} 
\label{fig:tan_rad_plane}
\end{center}
\end{figure}

We refer to the rotor path that results for $\Delta p= 0$ as the \textit{reference path} and only analyze deviations from this reference path from here on. We distinguish a radial deviation $s_{\text{R}}$ and a tangential deviation $s_{\text{T}}$ from the reference path as depicted in Figure~\ref{fig:tan_rad_plane}. It is convenient to define a rotating coordinate system with one axis representing $s_{\text{R}}$ and the other axis representing $s_{\text{T}}$ at each rotor end. For both rotor ends, the origin of the coordinate system is fixed to the center axis of the rotor. Figure~\ref{fig:tan_rad_plane} shows the rotating coordinate system and the two deviations $s_{\text{R}}$ and $s_{\text{T}}$ for two angles $\varphi_0$ and $\varphi_1$. We refer to this coordinate system as \textit{RT-coordinate system}.

The origin of the RT-coordinate system moves on the reference path of the rotor and rotates with the angular velocity $\dot{\varphi}$. 

The following analyses are carried out in the RT-coordinate system unless stated otherwise. We explain how to measure the rotor reference path and the determination of $K$ in section \ref{sec41}.

\section{Modeling backflow and pulsation in a PCP} \label{sec3}
The differential pressure across the PCP affects the tangential and the radial deviation of the rotor from its reference path in distinct manners. The radial deviation results in an almost constant tilt of the rotor with respect to the radial direction. The tangential deviation results in a periodic tilting of the rotor. This was observed and modeled successfully by \cite{mueller_2019} and is valid for the operating range of the pump. 
We claim the radial deviation governs the backflow, while the tangential deviation governs the pressure pulsation. 

\subsection{Backflow-conjecture} \label{sec31}
\begin{figure}
\begin{center}
\includegraphics[width=\columnwidth]{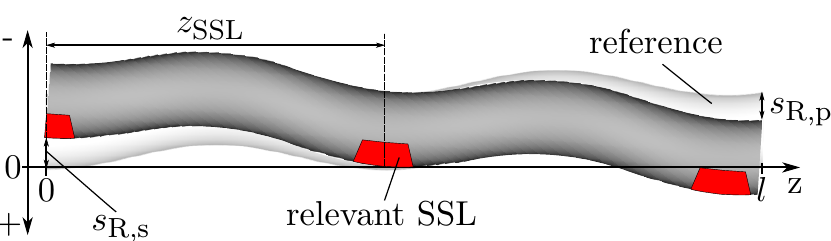}   
\caption{Correlation of radial deviations $s_{\small \textrm{R,s}}$ and $s_{\small \textrm{R,p}}$ and the location of the relevant SSL.} 
\label{fig:gapheight}
\end{center}
\end{figure}

In Section \ref{sec22} we showed that $\Delta p \not= 0$ holds for two SSLs. We will show in Section~\ref{sec42} that the shape of the SSL closest to the suction side degenerates most when differential pressure is applied.  
Thus, we analyze the shape of the SSL that is exposed to $\Delta p \not= 0$ and closest to the suction side. We refer to this specific SSL as the \textit{relevant SSL} (see Figure~\ref{fig:gapheight}). We assume that the backflow of the PCP mainly depends on the shape of the relevant SSL. This is referred to as the \textit{backflow-conjecture}. 

In the present section, we model the shape of the relevant SSL as a function of the measured radial deviation and the angle $\psi$. A simple model for the flow through the resulting gap is used to calculate the amount of backflow. The approach will be verified in Section~\ref{sec43} by comparing measured and calculated flow rates. 

By definition, the R-axis of the RT-coordinate system points towards the SSL (see Figures \ref{fig:rad_tan_coordinates} and \ref{fig:tan_rad_plane}). Thus, we expect that the shape of the SSL is strongly influenced by the radial deviation of the rotor from its reference path. 
The shape of the relevant SSL also depends on its distance to the suction side $z_{\small \textrm{SSL}}$. Figure~\ref{fig:gapheight} illustrates the correlation of the radial deviations to the location of the relevant SSL. $z_{\small \textrm{SSL}}$ can be described as a function of $\psi$ by
\begin{equation} \label{eq:SSL_location}
z_{\small \textrm{SSL}}=l - P_S\left(1- \frac{\mathrm{mod}(\psi- \theta, \pi)}{2 \pi} \right) \, ,
\end{equation}
with rotor length $l$, stator pitch $P_S$ and stator orientation $\theta$ (see \cite{mueller_2020} for additional information on why $\theta$ must be taken into account).

\begin{figure}
\begin{center}
\includegraphics[width=5.4cm]{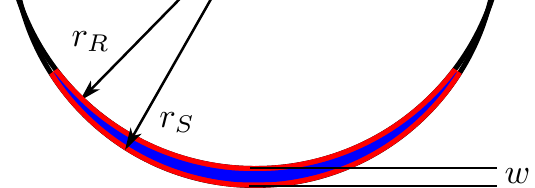}  
\caption{Resulting rotor-stator gap (area $A$ in blue, perimeter $U$ in red).} 
\label{fig:w}
\end{center}
\end{figure}

Rotor tilt leads to a vanishing contact between rotor and stator. The resulting gap height $w$ of the relevant SSL can be calculated from 
\begin{equation} \label{eq:gap_height}
w=\frac{s_{\small \textrm{R,p}}-s_{\small \textrm{R,s}}}{l} \cdot z_\textrm{\small SSL}+s_{\small \textrm{R,s}} \, ,
\end{equation}
with radial deviation at the pressure and suction side $s_{\small \textrm{R,p}}$ and $s_{\small \textrm{R,s}}$, respectively. Figure~\ref{fig:w} sketches the resulting gap at the relevant SSL. Because the radii of the rotor $r_R$ and the stator $r_S$ are known, $w$ can be used to calculate the gap area $A$ and the perimeter $U$. The intersection of rotor and stator (resulting from $r_R > r_S$) is neglected in \eqref{eq:gap_height}, so that any $w > 0$ leads to a breaking sealing. 
Note that for the specific case shown in Figure~\ref{fig:gapheight}, $w$ is close to zero, since the height at $z_{\small \textrm{SSL}}$ is close to the reference. Equation \eqref{eq:gap_height} reveals that $w$ is a function of $s_{\small \textrm{R,p}}$, $s_{\small \textrm{R,s}}$ and $\psi$. Consequently, $w$ varies during operation, even if the pump is operated at a fixed speed and differential pressure. However, our investigations revealed that it is sufficient to use a mean gap height $w_\text{{mean}}$, calculated over several rotations for the calculation of the backflow. This significantly reduces the computational effort.

The flow through the gap is assumed to be turbulent since our calculations yield high Reynolds numbers. 
We apply a model for the pressure loss in a pipe flow and use the hydraulic diameter $d_h=\frac{4A}{U}$ to calculate the pressure loss due to turbulences over the gap \cite{spurk_2020}. The pressure loss reads
\begin{equation} \label{eq:pressure_loss_turbulent}
\Delta p = \frac{\rho}{2} \cdot \left(\frac{Q_\textrm{b}}{A}\right)^2 \cdot \frac{L}{d_h} \cdot \lambda_t \, , 
\end{equation}
with differential pressure $\Delta p$, fluid density $\rho$, backflow $Q_\textrm{b}$, gap length $L$, and pressure loss coefficient $\lambda_t$. $\lambda_t$ can be described by the Blasius Equation $\lambda_t=\frac{0.3164}{(Re_{d_h})^{0.25}}$, with $Re_{d_h}=\frac{Q_\textrm{b} \cdot d_h \cdot \rho}{\eta \cdot A}$, where $\eta$ denotes the viscosity of the fluid \cite{hongqing_2018}. Inserting the Blasius Equation in (\ref{eq:pressure_loss_turbulent}) and solving it for $Q_\textrm{b}$ yields
\begin{equation} \label{eq:Q_L}
Q_\textrm{b}=A \cdot \left( \frac{2 \cdot \Delta p \cdot  d_h ^{1.25}}{\rho ^{0.75} \cdot L \cdot 0.3164 \cdot \eta^{0.25}} \right)^{\frac{1}{1.75}} \, .
\end{equation}
Equation \eqref{eq:Q_L} describes the backflow inside the PCP across the relevant SSL taking into account the geometry of the resulting gap due to the radial deviation of the rotor from its reference path. 

The calculated flow rate of the pump is determined by
\begin{equation} \label{eq:Q_cal}
Q_\textrm{cal}=Q_\textrm{i}-Q_\textrm{b} \, ,
\end{equation}
where $Q_\textrm{i}$ is the ideal flow rate of the pump.

\subsection{Pulsation-conjecture} \label{sec32}
\begin{figure}
\begin{center}
\includegraphics[width=\columnwidth]{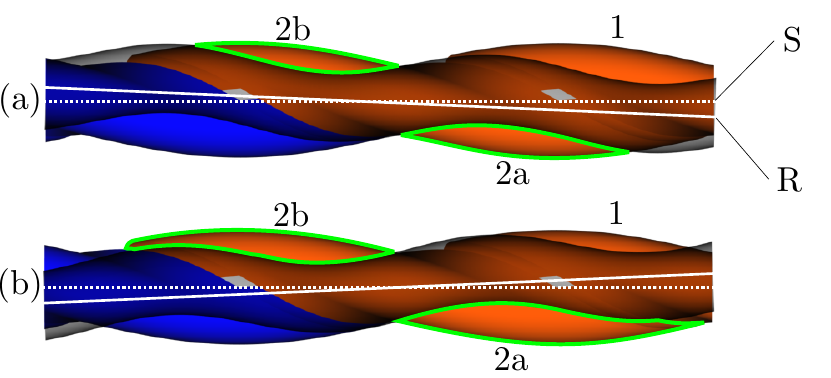}    
\caption{Increased cavity volume due to abrupt rotor tilt in tangential direction evident from the comparison of the stator center axis (S) and the rotor center axis of the rotor (R).} 
\label{fig:pulsation_tilt}
\end{center}
\end{figure}

Our second conjecture states that the pulsation in differential pressure, which can be observed during the operation of a PCP, mainly depends on the periodic tilting motion of the rotor in tangential direction. 
We use
\begin{equation} \label{eq:tilting_motion}
\tau(t)=s_{\small \textrm{T,s}}(t) - s_{\small \textrm{T,p}}(t)\, . 
\end{equation}
as a measure for the periodic tilting motion in tangential direction. We will show in Section~\ref{sec42} that $\tau$ abruptly changes twice per rotor rotation. Figure~\ref{fig:pulsation_tilt} sketches the position of the rotor before (a) and after (b) this abrupt tilting.  The abrupt change in $\tau$ leads to a quick volume increase in the most recently opened cavity 2 (highlighted by the green outline in Figure~\ref{fig:pulsation_tilt}). The quick increase in volume leads to an abrupt pressure drop on the pressure side of the pump.

\begin{figure}
\begin{center}
\includegraphics[width=\columnwidth]{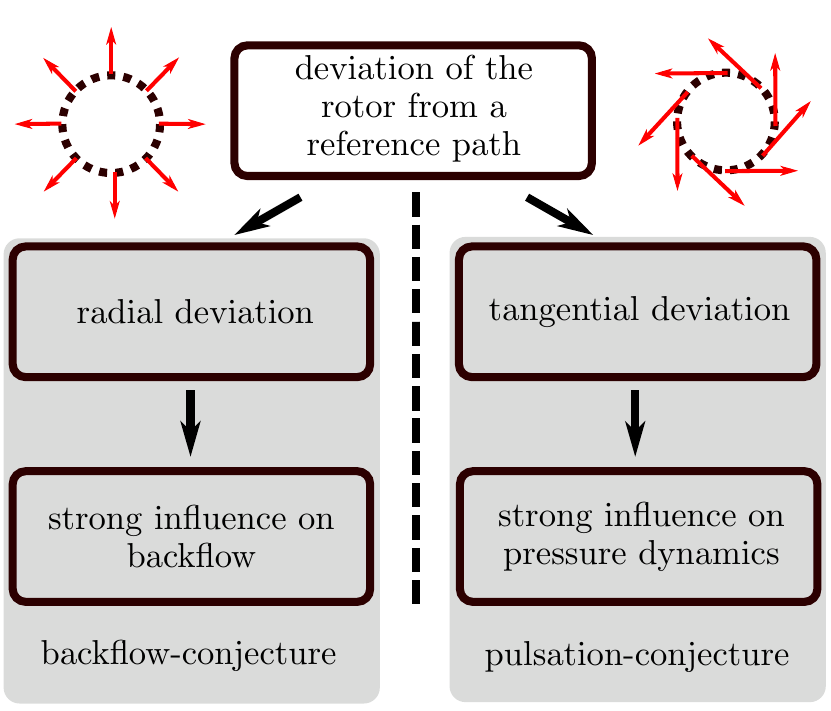}    
\caption{Effects of the deviation of the rotor from its reference path on the hydraulic conditions.} 
\label{fig:Movement_split}
\end{center}
\end{figure}

This conjecture will be verified in Section \ref{sec44} by comparing the measured periodic rotor tilt to the measured pressure pulsation on the pressure side. Figure~\ref{fig:Movement_split} summarizes our two conjectures.

\section{Experimental verification} \label{sec4}
This section presents the laboratory test setup and verifies the claims made so far experimentally.
\subsection{Laboratory test setup} \label{sec41}
\begin{figure}
\begin{center}
\includegraphics[width=\columnwidth]{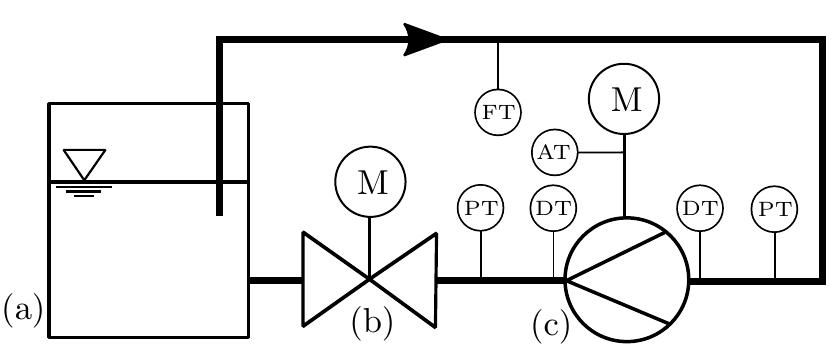}    
\caption{Scheme of the laboratory test setup with the container (a), the control valve (b), the progressing cavity pump (c), the flow transmitter FT, the angular transmitter AT, the pressure transmitters PT and the distance transmitters DT.} 
\label{fig:test_setup}
\end{center}
\end{figure}
The experimental data used in this paper is measured with the laboratory test setup sketched in Figure~\ref{fig:test_setup}. The progressing cavity pump\footnote{10-6L, manufactured by Seepex GmbH} (c) pumps water from the container (a) through the control valve (b) and back into the container. The pump is driven with a variable frequency drive. Various differential pressures can be set with the control valve. 

The angle $\psi$, the rotational speed $n$, the flow rate $Q$, and the pressure and suction side pressure are measured as indicated in Figure~\ref{fig:test_setup}.

Distance transmitters (8 at each side of the rotor) measure the distance between each end of the rotor and the pump housing during the operation (see \cite{mueller_2017} for details). This allows to determine $\left(x_s(t), y_s(t) \right)'$ and $\left(x_p(t), y_p(t)\right)'$ (see Section~\ref{sec23}) as a function of time \cite{mueller_2019}, thus measuring the rotor reference path as well as the deviations due to differential pressure. We initialize $\mathcal{S}$ and $\mathcal{R}$ with manufacturing data and manually determine $K$ in \eqref{eq:delta_d} so that no sealing in $\mathcal{C}$ degenerates when the pump is operated with zero differential pressure (see assumption in Section \ref{sec22}).

All signals are measured with a sample time of 1\,ms.

\subsection{Degeneration of the sealings due to differential pressure} \label{sec42}
\begin{figure}
\begin{center}
\includegraphics[width=\columnwidth]{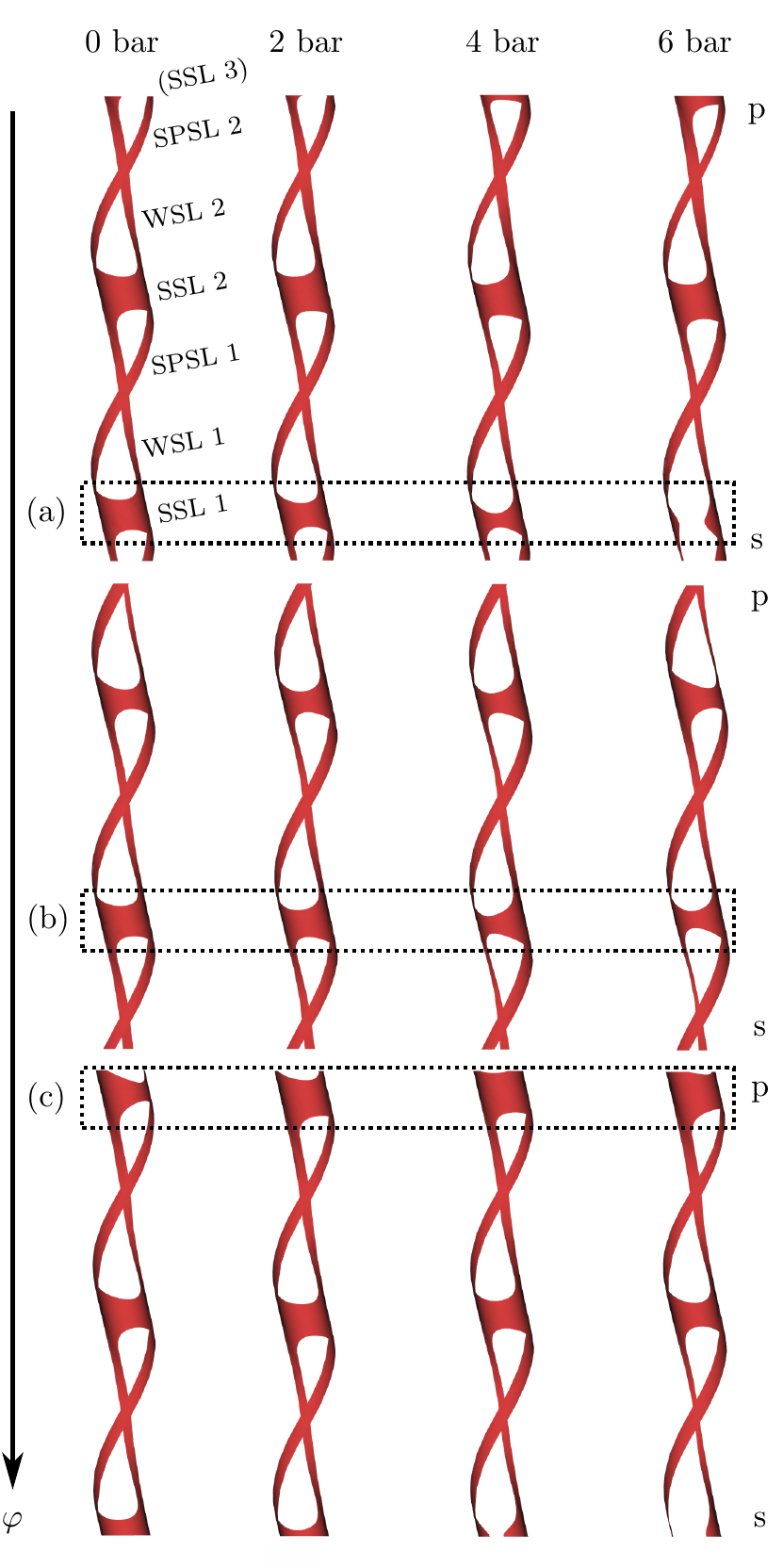}   
\caption{Reconstructed sealings $\mathcal{C}$ for various differential pressures and rotational angles $\varphi$. The pump was operated at 100\,rpm. Suction side indicated by s, pressure side indicated by p. Dashed boxes highlight SSLs at various locations.} 
\label{fig:SSL_quality}
\end{center}
\end{figure}

We analyze the shape of the sealings as a function of differential pressure. Representative results obtained with the geometric 3D model\footnote{We implement the geometric 3D model in the CAD Software FreeCAD and use its python interface to place the geometric representation of the stator as well as the rotor according to measured values.} and the correction factor $K \approx 1.014$ are shown in Figure~\ref{fig:SSL_quality}.

Figure~\ref{fig:SSL_quality} reveals that the SPSL and the WSL are almost insensitive to changes in the movement of the rotor due to increasing differential pressure. In contrast, the SSL significantly depends on the differential pressure. The contact area of the SSL 1 decreases with increasing differential pressure (see Figure~\ref{fig:SSL_quality} (a)). With increasing $\varphi$, SSL 1 moves towards the pressure side (see Figure~\ref{fig:SSL_quality} (b)). At this location, the area of the considered SSL only weakly depends on the differential pressure. The pressure dependence is reversed close to the pressure side, which leads to an increasing contact area of the considered SSL with increasing differential pressure (see Figure~\ref{fig:SSL_quality} (c)). 

The results in Figure~\ref{fig:SSL_quality} suggest that the amount of backflow mainly depends on the shape of the SSL closest to the suction side, because any SSL closer to the pressure side exhibits a bigger contact area. 

Additionally, both deviations of the rotor from its reference path due to applied differential pressure were calculated for various rotational speeds and differential pressures ranging from 0 to 4 bar, varied by adjusting the control valve (b) in Figure \ref{fig:test_setup}.

\begin{figure}
\begin{center}
\includegraphics[width=\columnwidth]{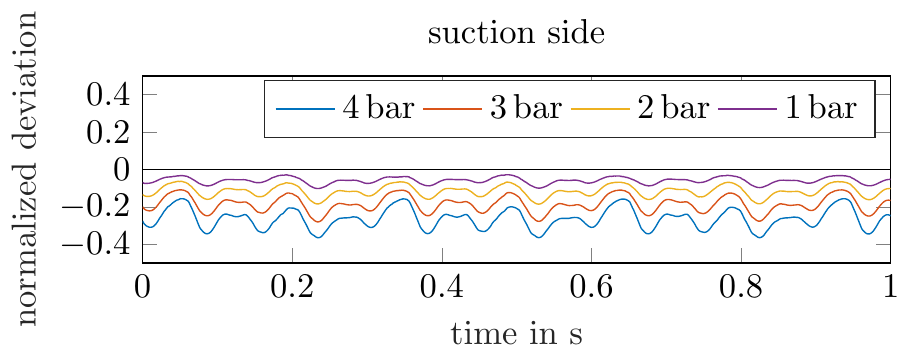}   
\includegraphics[width=\columnwidth]{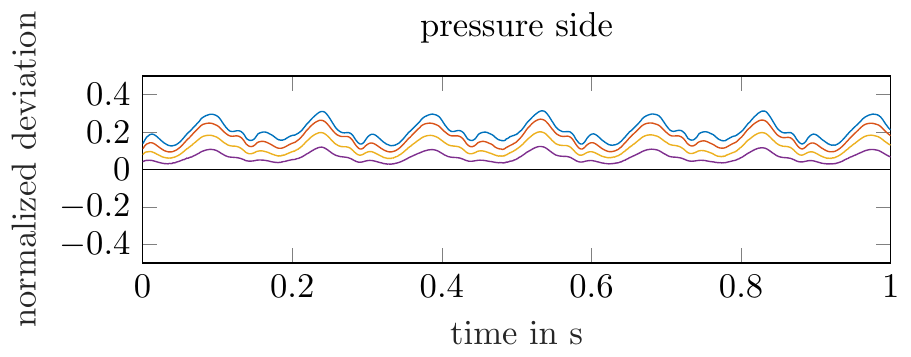}
\caption{Radial deviation of the rotor from its reference path for 200 rpm and various differential pressures.} 
\label{fig:radial_deviation}
\end{center}
\vspace{-0.4cm}
\end{figure}

Figure~\ref{fig:radial_deviation} shows the radial deviation of the rotor for the suction side and the pressure side for 200\,rpm and various differential pressures. All deviations have been normalized by dividing them by the maximum value $s_{\textrm{max}}$ that occurs in the tangential deviation of the pressure side.\\

The tilt of the rotor in radial direction, which depends on the applied differential pressure, is shown in Figure~\ref{fig:radial_deviation}.

\begin{figure}
\begin{center}
\includegraphics[width=\columnwidth]{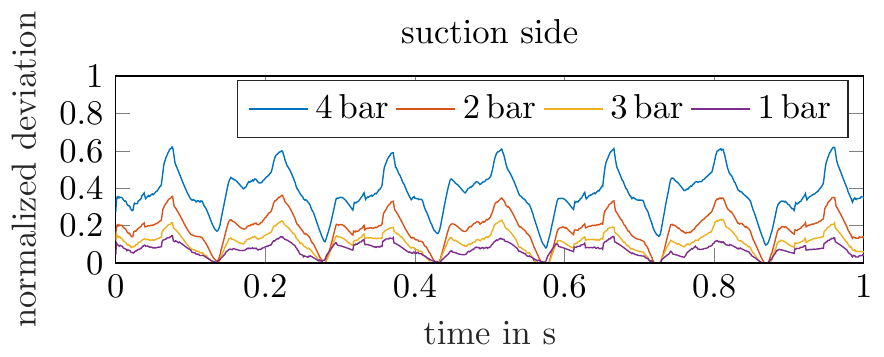}    
\includegraphics[width=\columnwidth]{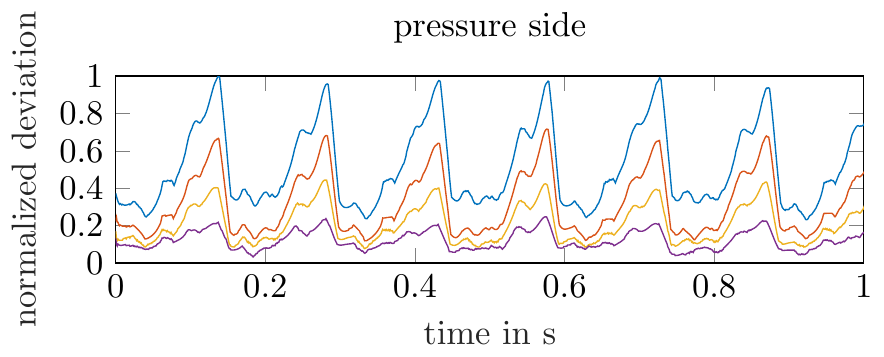}
\caption{Tangential deviation of the rotor from its reference path for 200 rpm and various differential pressures.} 
\label{fig:tangential_deviation}
\end{center}
\vspace{-0.2cm}
\end{figure}

The normalized tangential deviation of the rotor from its reference path is depicted in Figure~\ref{fig:tangential_deviation} for the same points of operation. 
\begin{figure}
\begin{center}
\includegraphics[width=\columnwidth]{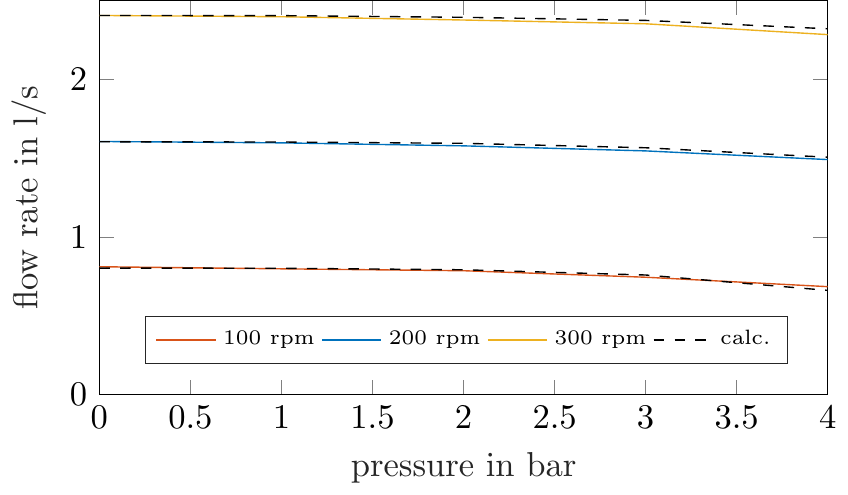}  
\caption{Comparison of measured and calculated flow rates for various rotational speeds and differential pressures.} 
\label{fig:flowrate}
\end{center}
\end{figure}
Figure~\ref{fig:tangential_deviation} shows the periodic tilting motion of the rotor addressed in Section~\ref{sec32}.

\subsection{Verification of the backflow-conjecture} \label{sec43}
We compute the backflow according to the approach outlined in Section~\ref{sec31}. The calculation of $Q_\textrm{b}$ uses the mean gap height $w_\text{{mean}}$ and is thus only performed once for each combination of speed and differential pressure. The specific value for the gap length $L$ in \eqref{eq:Q_L} was found by the comparison between the measured and calculated flow rate for one arbitrary point of operation with non-zero differential pressure and kept constant subsequently. It was found to be $L=2.5 mm$, which is physically reasonable and is in a similar magnitude as the value reported in \cite{pessoa_2009}.

Figure~\ref{fig:flowrate} shows the comparison of the calculated to the measured flow rates for various speeds and differential pressures. The error between the calculated and the measured flow rate is below 3.5\,\%. 
The largest absolute error between measured and calculated backflow occurs at 300 rpm and 4 bar differential pressure and amounts to 0.037 l/s.

The proposed approach reproduces the measured flow rates with reasonable accuracy. This corroborates the proposed backflow-conjecture.

\subsection{Verification of the pulsation-conjecture} \label{sec44}
\begin{figure}
\begin{center}
\includegraphics[width=\columnwidth]{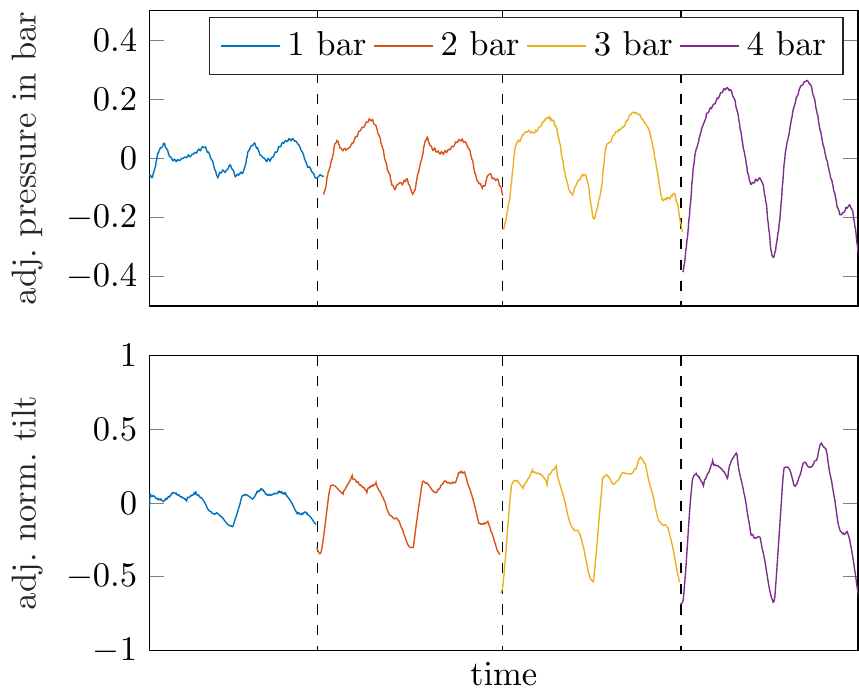}    
\caption{Time series of pressure $p_\text{p}$ and rotor tilt $\tau$ for 200 rpm and various differential pressures.} 
\label{fig:pulsation}
\end{center}
\end{figure}

Figure~\ref{fig:pulsation} shows time series of $\tau$ and $p_\text{p}$ for various points of operation.

The time series for each point of operation is depicted for one revolution of the rotor. All time series in Figure~\ref{fig:pulsation} are adjusted by subtracting their mean value for easier comparison in a joint figure. The strong correlation of $\tau$ to $p_\text{p}$, evident in Figure~\ref{fig:pulsation}, corroborates the pulsation-conjecture.

\section{Conclusion} \label{sec5}
In this contribution, a novel approach to modeling the backflow of a PCP was proposed. The model takes the actual movement of the rotor due to its interaction with the deformable stator into account. The deviation of the rotor from a reference path inside the elastomer stator was divided into two components. We showed that the radial deviation can be used to model the backflow, whereas the tangential deviation mainly affects pressure dynamics. 
In contrast to existing approaches, the proposed model takes into account information regarding the actual, non-ideal movement of the rotor. 

\section*{Acknowledgments}
Funding by Ministerium f\"ur Wirtschaft, Innovation, Digitalisierung und Energie des Landes Nordrhein-Westfalen is greatfully acknowledged.
\begin{center}
\includegraphics[width=\columnwidth]{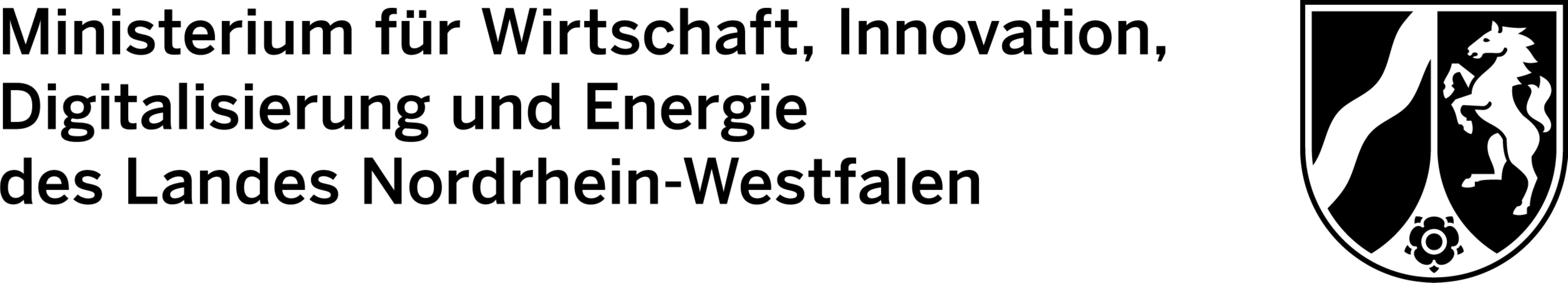}   
\end{center}
\small
\bibliographystyle{plain}
\bibliography{jpse2020}  
The formal publication of this preprint can be found via https://doi.org/10.1016/j.petrol.2021.108402

\end{document}